\documentstyle[12pt,epsf]{article}

\newlength{\dinwidth}
\newlength{\dinmargin}
\newcommand{\be}{\begin{equation}}
\newcommand{\ee}{\end{equation}}
\newcommand{\bc}{\begin{center}}
\newcommand{\ec}{\end{center}}
\newcommand{\ba}{\begin{array}}
\newcommand{\ea}{\end{array}}
\newcommand{\bd}{\begin{displaymath}}
\newcommand{\ed}{\end{displaymath}}
\newcommand{\bea}{\begin{eqnarray}}
\newcommand{\eea}{\end{eqnarray}}
\newcommand{\beaa}{\begin{eqnarray*}}
\newcommand{\eeaa}{\end{eqnarray*}}
\newcommand{\btab}{\begin{tabular}}
\newcommand{\etab}{\end{tabular}}
\newcommand{\bfig}{\begin{figure}}
\newcommand{\efig}{\end{figure}}
\newcommand{\eq}[1]{\mbox{(\ref{#1})}}
\newcommand{\fig}[1]{\mbox{fig.\ \ref{#1}}}

\newcommand{\eg}{\mbox{e.g.}}
\newcommand{\ie}{\mbox{i.e.}}

\newcommand{\Z}{\mbox{\rm\bf Z}}
\newcommand{\z}{\mbox{\rm\scriptsize\bf Z}}

\newcommand{\slas}[1]{#1\!\!\!/}

\newcommand{\abs}[1]{|#1|}
\newcommand{\inv}[1]{\frac{1}{#1}}

\newcommand{\mbbm}[1]{{\mbox{\boldmath $#1$}}}

\newcommand{\preprint}[2]{\baselineskip15pt
  \begin{flushright} \begin{tabular}{l}
      #1 \\[2mm] #2
  \end{tabular} \end{flushright}
}
\renewcommand{\title}[1]{\begin{center}\parbox{12cm}
  {\baselineskip30pt\centering{{\Large{\bf #1}}}}\end{center}
}
\newcommand{\cpc}[3] {Comp.\ Phys.\ Comm.          {#1} {(#2)} {#3}}

\newcommand{\prd}[3] {Phys.\ Rev.\ D               {#1} {(#2)} {#3}}

\newcommand{\Res}[2]{
  \begin{array}[t]{c}
    \mbox{Res} \\ [-1.1ex]{\scriptscriptstyle #1}
  \end{array} \! \! \left\{#2 \right\}
}
\newcommand{\Math}{{\em Mathematica\/}}
\newcommand{\FS}{{\em FreSum\/}}

\begin{document}
%
%%%%%%%%%%  AUTHOR, TITLE, PREPRINT #, DATE  & ABSTRACT  %%%%%%%%%%%%
%
\def\thefootnote{\fnsymbol{footnote}}
\baselineskip18pt
\thispagestyle{empty}
\preprint{FTUAM-93/19 \\ hep-ph/9311210}{October 29, 1993}
\vspace*{1.5cm}
\title{EVALUATING SUMS OVER THE MATSUBARA
  FREQUENCIES\footnotemark}\vspace*{1in}
\footnotetext{This work was partially supported by the CICyT (Spain)
  under contract AEN/93/673.}
\baselineskip18pt
\begin{center}
  Agustin {\sc NIETO}\footnote{Address after January 1, 1994:
    Department of Physics and Astronomy, Northwestern University,
    Evanston, Illinois 60208, USA.}  \\
    {\small  Departamento de Fisica Teorica, C-XI,} \\[-2mm]
    {\small  Universidad Autonoma de Madrid,
             E-28049 Madrid, Spain                      } \\[-2mm]
    {\small E-mail: {\tt anieto@madriz1.ft.uam.es}}
\end{center}
\vspace*{.7in}
\baselineskip24pt
\indent
%
%  ABSTRACT
%
Perturbative calculations in field theory at finite temperature
involve sums over the Matsubara frequencies. Besides the usual
difficulties that appear in perturbative computations, these sums give
rise to some new obstacles that are carefully analized here.
I present a fast and realible recipe to work out sums over the
Matsubara frequencies. As this algorithm leads
to deal with very cumbersome algebraic expressions, it has been
written for computers by using the symbolic manipulation program \Math.
It is also shown this algorithm to be self-consistent when it is applied to
more than one loop computations.
%
%%%%%%%%%%%%%%%%%%%%%%%%%%%%%%%%%%%%%%%%%%%%%%%%%%%%%%%%%%%%%%%%%%%%%
%
\baselineskip18pt
\setcounter{page}{0}
\def\thefootnote{\arabic{footnote}}
\setcounter{footnote}{0}
\newpage
%
%%%%%%%%%%%%%%%%%%%%%  TEXT  BEGINS  HERE  %%%%%%%%%%%%%%%%%%%%%%%%%%
%
A system of particles at temperature $T=1/\beta$ is completly
described by its partition function. From it, it is very easy to get the
other physical magnitudes, \eg\ pressure, energy, entropy, etc.  So, one
of the main tasks of a statistical field theory is to develop methods to
obtain the partition function. This can be acomplished by means of a
perturbative expansion in the coupling constant. Following this idea, one
finds that the partition function is represented by the
one-particle-irreducible Feynman diagrams without external legs (bubble
diagrams). This is just a simple example to point out the
importance of perturbative methods in finite temperature field theory,
which have been widely and sucessfully applied to different purposes as
the QCD deconfinement.

At finite temperature, together with the
 usual difficulties that a perturbative computation
involves (integration over loops, regularization and renormalization),
some others arise because of the sums over the
Matsubara frequencies. The time-like component of the momenta is quantized
because the definition of the partition function requires the fields
to be periodic in
time (fermions are anti-periodic). Its allowed values are the so-called
Matsubara frequencies: $i\,2n\pi/\beta$ for bosons and $i(2n+1)\pi/\beta$ for
fermions ($n\in\Z$). This makes the integration over that component change into
a sum over frequencies~\cite{Kap89:1}.

We will focus on evaluating sums over the
Matsubara frequencies. The method that I will describe in what follows is
well known but I think is worth noticing its self-consistency on
evaluating each sum over the Matsubara frequencies of more-than-one
loop diagrams.
Also, a computer program which provides a fast and realible way of
performing these sums is presented. Finally, some examples are worked
out by using such program.

Let us sum a function $f$ over the set of frequencies
$k^0=i \omega_n= i\,2n\pi/\beta$
\be
  \frac{1}{\beta} \sum_{n \in \z} f(k^0). \label{sumfr}
\ee
If $f$ does not have any singularity along the imaginary axis, we will
multiply it by a function with simple poles and residue one at $k^0=i
\omega_n$, \eg, $1/(2 \pi i) (\beta/2) \coth (k^0 \beta/2)$. So,
\eq{sumfr} is written
\be
  \frac{1}{2 \pi i} \oint_{C} dz \; f(z) u_{\beta}(z), \label{oint}
\ee
where $u_{\beta} (z) \equiv \frac{1}{2} \coth (\beta z/2)$ and $C$ is the
contour in \fig{conto}(a). This contour can be
deformed into the one shown in \fig{conto}(b) and, therefore,
eq.~\eq{oint} can be expressed as
\be
  - \frac{1}{2 \pi i} \int_{-i\infty-\epsilon}^{i\infty-\epsilon}
    dz \; f(z) u_{\beta}(z)
    + \frac{1}{2 \pi i} \int_{-i\infty+\epsilon}^{i\infty+\epsilon}
    dz \; f(z) u_{\beta}(z)
    .  \label{intes}
\ee
Using the contours in \fig{conto}(c), we can write
the integrals in \eq{intes} as
\bd
  \ba{ccc}
     \displaystyle{\int_{-i\infty-\epsilon}^{i\infty-\epsilon}
     = \oint_{C_1} - \int_{\Gamma_1}}
     & \mbox{and} &
     \displaystyle{\int_{-i\infty+\epsilon}^{i\infty+\epsilon}
     = \int_{\Gamma_2} - \oint_{C_2}}.
  \ea
\ed
Now, if $f(z) u_\beta(z)$ goes fast enough to zero as
$|z|\rightarrow\infty$, the integrals at infinity will vanish.
In this case, we conclude
\be
  \frac{1}{\beta} \sum_{n \in \z}
  f\left( k^0 = i\,2n\pi/\beta \right) = - \sum_{a}
  \Res {z=z_a}{ f(z) u_{\beta}(z) }
  ,  \label{boson}
\ee
where $z_a$ are the poles of $f(z)$.

A similar result can be obtained for fermions at zero chemical
potential
\be
  \frac{1}{\beta} \sum_{n \in \z}
  f\left( k^0 = i(2n+1)\pi/\beta \right) = - \sum_{a}
  \Res {z=z_a}{ f(z) v_{\beta}(z) }
  ,  \label{fermion}
\ee
where $v_{\beta}(z)\equiv\frac{1}{2}\tanh(\beta z/2)$.

Considering the dependence on the momenta in a Feynman
diagram we can go further in our analysis. The contribution of an
$L$-loop Feynman diagram
with $4$-momentum $k_j=(k^0_j,\mbbm{k}_j)$ flowing around each loop
($j=1,\dots,L$) can be written as
\be
  \int_{\mbbm{k}_1}\dots\int_{\mbbm{k}_L}
  \inv{\beta}\sum_{k_1^0}\dots\inv{\beta}\sum_{k_L^0}
  f(k_1^0,\dots,k_L^0); \label{FD}
\ee
where $k_j^0=i\,\omega_{n_j}$ is the corresponding Matsubara frequency,
$f$ also depends on $\mbbm{k}_1\dots\mbbm{k}_L$ and in general has the
form
\be
  f(k_1^0,\dots,k_L^0)=\sum{
  h(k_1^0,\dots,k_L^0)\prod_{R=1}^{n}
  D^{\gamma_R}\left(\sum_i \sigma_i^R k_i^0+a^R,b^R\right) };\label{efe}
\ee
The $D$-function is defined as
\be
  D(x,y)=\inv{x^2-y^2}; \label{defu}
\ee
it is a non-analytical function that arises by noticing that, with
$E_k^2\equiv\mbbm{k}^2+m^2$, $D(k^0,E_k)$ is the propagator
for a scalar field, $(\slas{k}+m)D(k^0,E_k)$ is the one for
fermions, $-g_{\mu\nu}D(k^0,\abs{\mbbm{k}})+(1-\xi)k_\mu k_\nu
D^2(k^0,\abs{\mbbm{k}})$ is the one for photons and similarly for other fields.
The function $h$ is analytic on $k_1^0,\dots,k_L^0$;
$\sigma_i^R=\{+1,0,-1\}$; $a^R$ and $b^R$ do not depend on $k_i^0$ and the
exponent $\gamma_R$ is an integer. The global sum means that $f$ may be a
sum of functions with these properties.

An interesting property of the Feynman diagrams is that if
$\sigma_i^R$, $\sigma_j^R$, $\sigma_i^{R'}$ and $\sigma_j^{R'}\neq 0$
\be
  \sigma_i^{R}\sigma_j^{R}=\sigma_i^{R'}\sigma_j^{R'}\hspace{1cm}
  \forall R\neq R'.
\ee
In other words, if a propagator in the diagram carries
momentum $k-p+\Sigma_1$, there may exist other propagators with momentum
$-k+p+\Sigma_2$, but not $k+p+\Sigma_3$ nor $-k+p+\Sigma_4$, \ie\ the
relative sign of two momenta, \eg\ $k$ and $p$, is constant.

A detailed and straightforward analysis shows that the result of
performing a sum over, let us say, the $j$th Matsubara frequency of a
function that verifies the previous properties is another function that
verifies the same properties. So, if we were able to find a method to
evaluate such a sum, we would be able to perform iteratively every sum.

Let us use eq.~\eq{boson} to compute
\be
  \inv{\beta}\sum_{k_L^0} f(k_1^0,\dots,k_L^0)
\ee
for a scalar field. Using the~\eq{efe} and~\eq{defu} one finds that
when $\sigma^R_L\neq 0$ each $D$-function has
two poles of order $\gamma_R$ at
\be
  k_L^0=-\sigma_L^R\left(\sum_{i\neq L}\sigma_i^R k_i^0+a^R_0 \pm b^R
  \right)\equiv z_{\pm}^R.
\ee
It is worth stressing that the function $f$ has no other poles but the ones
given by the $D$-functions. So,
\bea
  \inv{\beta}\sum_{k^0_L}f(k_1^0,\dots,k_L^0)&=&-\sum_{R':\sigma^{R'}\neq 0}
  \left\{\Res{z=z_+^{R'}}{f u_\beta(z)} + \Res{z=z_-^{R'}}{f u_\beta(z)}
  \right\} \nonumber \\
  &=& -\sum_{R':\sigma^{R'}\neq 0}\Res{z=z_+^{R'}}{f u_\beta(z)} +
     (b^{R'}\rightarrow-b^{R'}).
\eea
Where,
\be
  \Res{z=z_+^{R}}{f u_\beta(z)}=\inv{(\gamma_R-1)!}
  \lim_{z\rightarrow z_+^R}\frac{d^{\gamma_R-1}}{dz^{\gamma_R-1}}
  \left\{(z-z_+^R)^{\gamma_R}f(k_1^0,\dots,k_{L-1}^0,z)u_\beta(z)\right\}.
  \label{resbos}
\ee
The result for fermions would have been obtained simply by changing
$u_\beta(z)$ into $v_\beta(z)$. Equation~\eq{resbos} and its
fermionic equivalent are all we need to perform a sum over a single
frequency; additional sums can similarly be computed by following
the previous analysis.

So far, we have not considered whether evaluating these sums is completly
sensible without mentioning the possible divergences. We just evaluated
them by using a regularization method; so they do not indeed diverge. The
expected divergences that will arise when calculating the space-like
integrals should be handled by the usual renormalization schemes.

Summarizing, the recipe
consists on identifying the poles of the $D$-functions and evaluating the
residues at these poles. Then every residue is summed. The result is
another function on the remaining momenta.

However, this apparently simple scheme cannot be easily acomplished for
diagrams with a large number of propagators and loops. The amount
of algebra nedeed when dealing with physically interesting cases
become an akward task to be performed by hand. Computers, though, have
shown to be good at doing cumbersome and repetitive tasks.

Symbolic manipulation programs have been developed to handle algebra.
Some of them are intended to perform perturbative calculations in
physics, \eg\ Schoonship~\cite{VeW91:1}, REDUCE~\cite{Hea87:1} and
FORM~\cite{Ver89:1}. On the other hand, there are general purpose
programs, like MAPLE~\cite{Wat88:1} or \Math~\cite{Wol88:1}, that provide
functions that allow one to do additional manipulations on the results,
\eg\ integrate over the space-like momenta, plot results, etc. In this
context, I present here a \Math\ package that evaluates the
sums over the Matsubara frequencies and that, therefore, helps to
calculate Feynman diagrams at finite temperature.
I have called this package \FS\footnote{This package is available at
the directory {\tt pub/math/paquetes} of the Anonymous FTP site
{\tt ftp.uam.es}.}.

Let us compute the one-loop correction to the self-enegy of a scalar
field interacting with itself through a $\lambda\phi^4$ potential. It will be
(see \fig{fg:421})
\bea
  \Pi_{\beta} & = &\frac{\lambda}{2} \int \frac{d^3k}{(2 \pi)^3}
  \frac{1}{\beta} \sum_{k^0}\frac{1}{(k^0)^{2}-E_{k}^{2}}
  \nonumber \\
  &=&\frac{\lambda}{4\pi^2} \int_0^{\infty} dk\, k^2
  \inv{\beta} \sum_{k^0}D(k^0, E_k)
  \label{eq:421}
\eea
Once \Math\ is started, the following commands are needed to perform the
sum over the quantized time-like momentum:
\begin{verbatim}
        In[1]= <<fresum`
        In[2]= Even[k0];
        In[3]= f[k0_] = dd[k0,Ek,1];
        In[4]= result0 = FreSum[f[k0],k0]
\end{verbatim}
The first line instructs \Math\ to load the package \FS. The following
instruction defines $k^0\equiv i\,2n\pi/\beta$ as an even Matsubara
frequency (if we had considered a spinor field, it would have been an
odd Matsubara frequency: $i (2n+1)\pi/\beta$, and we would have used the
\FS\ function {\tt Odd[]}). The third line defines the function $f(k^0)$
that we want to sum; here {\tt dd[]} is the \Math\ $D$-function. Finally we
call {\tt result0} to the result of performing the sum over $k^0$; it is
\bd
  -\frac{u_{\beta}(E_k)}{E_k}.
\ed
Now, we can go further and take the massless limit
\begin{verbatim}
        In[5]= result1 = result0 /. Ek->k;
\end{verbatim}
We can put aside the dependence on the temperature with the
\FS\ functions {\tt VacuumPart[]} and {\tt MatterPart[]}; the vacuum part
is the $T\rightarrow 0$ limit of the argument, while the matter part is
the result of taking away the vacuum part from the argument.
\begin{verbatim}
        In[6]= result2 = MatterPart[result1]
        In[7]= result3 = result2 /. Nb[x_,bb] -> 1/(Exp[bb x]-1)
\end{verbatim}
we get $-N_{\beta}^B(k)/k$, where {\tt bb} is $\beta=1/T$,
$k=\abs{\mbbm{k}}$ and
\be
  N_{\beta}^B(z)\equiv u_{\beta}(z)-\inv{2}=\inv{\exp(\beta z)-1}
\ee
is the bosonic occupation number. If spinor fields are involved,
the result will be written by using
\be
  N_{\beta}^F(z)\equiv \inv{2}-v_{\beta}(z)=\inv{\exp(\beta z)+1}
\ee
which is the fermionic occupation number. The integration over the
3-momentum can be performed:
\begin{verbatim}
    In[8]= result4 = lambda k^2 result3 / (4 Pi^2)
    In[9]= result = Integrate[result4,{k,0,Infinity}]
\end{verbatim}
Finally, the result for the self-energy is
\be
  \Pi_{\beta}=-\frac{\lambda}{24\beta^2} \label{phi4}
\ee
which coincides which the one obtained in~\cite{Kap89:1}
and~\cite{DJT74:1} for this case.

Another interesting calculation that can easily be done with the help of \FS\
is the 1-loop photon self-energy with zero external momentum. The
$00$-component can be written in the Feynman gauge ($\xi=1$)
\be
  \Pi_\beta^{00}(0)=-4 e^2\int\! d^3k\;
  \inv{\beta}\sum_{k_0}(k_0^2+E_k^2)D^2(k_0,E_k).
\ee
The instructions
\begin{verbatim}
        In[1]= <<fresum`
        In[2]= Even[p0];
        In[3]= Odd[k0];
        In[4]= f[k0_] = (k0^2+Ek^2) dd[k0,Ek,2];
        In[5]= result0 = -4 ee2 FreSum[f[k0],k0]
        In[6]= result1 = 4 Pi k^2 result0 / (2 Pi)^3
        In[7]= result2 = Simplify[MatterPart[result1]]
\end{verbatim}
give rise to
\be
  \Pi_\beta^{00}(0)=-\frac{2e^2\beta}{\pi^2}\int_0^{\infty}dk\, k^2
  N_\beta^F(E_k)\left[N_\beta^F(E_k)-1\right]. \label{qed}
\ee
Note that the third argument of {\tt dd[]} is the power of the
corresponding $D$-function. The sixth instruction simulates the angular
integration. In this case, the vacuum part vanishes in the zero external
momentum limit due to the Lorentz invariance of the theory; so, here, the {\tt
MatterPart[]} function just changes {\tt Uo[]} into {\tt Nf[]}. Now, writing
$N_\beta^F(z)$ explicitly and performing the changes of variable
$k\beta\rightarrow k$ we find that the photon self-energy is
\be
  \Pi_\beta^{00}(0)=\frac{e^2}{3\beta^2}g(m\beta)
\ee
where the function
\be
  g(z)\equiv \frac{6}{\pi^2}\int_0^{\infty}dk\,
  \frac{k^2\exp\sqrt{k^2+z^2}}{\left(1+\exp\sqrt{k^2+z^2}\right)^2}
\ee
is shown in~\fig{gfun}. The following chain of instructions will produce such
result:
\begin{verbatim}
        In[8]= result3 = Simplify[result2 /. Nf[x_,bb]->1/(Exp[bb x]+1)]
        In[9]= result4[k_,z_] = Simplify[result3
          /. {bb->1, ee2->1,Ek->Sqrt[k^2+z^2]}]
        In[10]= g[z_] := 3 Integrate[result4[k,z],{k,0,Infinity}]
        In[11]= Plot[g[z], {z,0,10}]
\end{verbatim}
As $g(0)=1$, we recover the well known result in the massless
limit~\cite{Kap89:1,For80:1}
\be
  \Pi_\beta^{00}(0)=\frac{e^2}{3\beta^2}.
\ee

So far, perturbative calculations have
almost exclusively been done by using numerical methods. However, when
one is concerned with topics that require a non-limit behavior, additional
algebra arises and, therefore, symbolic manipulation
programs are required in order to use computers efficiently. This is the
context in which the package that we have described here will be a useful
tool.

We have seen that the evaluation of the sums over the Matsubara
frequencies can easily be accomplished. Two simple examples have
been worked out to show how the package works. A {\em NeXTstation},
spent less than 2 sec.\ to get~\eq{phi4}.
The photon self-energy~\eq{qed} calculation took 2.5 sec.,
while \fig{gfun} took $1,\!500$ sec.\ because the required
integrations have to be worked hard by \Math. We can learn a number of
facts from these computations; the sums over the Matsubara frequencies are
algebraically awkward, but they are performed systematically, no
matter the number of loops. One eventually gets a result that has
to be integrated over the 3-momenta; this turns out to be the main
obstacle to completly evaluate an involved diagram.
This task, when dealing with more-than-one loop diagrams, will
probably require numerical methods of integration to be achived.
\section*{Acknowledgments}
I am grateful to Enrique Alvarez for posing me the original work that gave rise
to this paper as a part of my Ph.D. thesis. Also, I want to thank
M.~A.~R.~Osorio and M.~A.~Vazquez-Mozo for fruitful discussions.

%%%%%%%%%%%%%%%%%%%%%%% TEXT  ENDS  HERE  %%%%%%%%%%%%%%%%%%%%%%%%%%%
%
%%%%%%%%%%%%%%%%%%%%%%%%%  REFERENCES  %%%%%%%%%%%%%%%%%%%%%%%%%%%%%%
%

%
%%%%%%%%%%%%%%%%%%%%%  TABLE  OF  FUNCTIONS  %%%%%%%%%%%%%%%%%%%%%%%%
%
\newpage
\bc {\Large\bf List of FreSum Objects.} \ec
\btab{ll}
  \hline\hline \\
  {\tt Fresum[f, k0]}    & Performs the sum of {\tt f} over the {\tt k0}
                           Matsubara frequency. \\
                         & $\inv{\beta}\sum_{k^0}f(k^0)$ \\ \\
  {\tt VacuumPart[expr]} & The $T=0$ limit of {\tt expr}. \\ \\
  {\tt MatterPart[expr]} & The result of taking away
                           {\tt VacuumPart[expr]} from {\tt expr}. \\ \\
  {\tt dd[x,y,N]}        & The \Math\ version of the {\tt N}th
                           power of the \\
                         & $D$-function: $D(x,y)=1/(x^2-y^2)$. \\ \\
  {\tt bb}               & The \Math\ version of $\beta=1/T$, the
                           inverse temperature. \\ \\
  {\tt Ue[z,bb]}         & The \Math\ version of
                           $u_\beta(z) = \inv{2}\coth(\beta z/2)$. \\ \\
  {\tt Uo[z,bb]}         & The \Math\ version of
                           $v_\beta(z) = \inv{2}\tanh(\beta z/2)$. \\ \\
  {\tt Nb[z,bb]}         & The bosonic occupation number:
                           $N_\beta^B(z)=1/(\exp(\beta z)-1)$. \\ \\
  {\tt Nf[z,bb]}         & The fermionic occupation number:
                           $N_\beta^F(z)=1/(\exp(\beta z)+1)$. \\ \\
  {\tt Even[x1,x2,...]}  & {\tt x1,x2,...} are defined as even Matsubara
                           frequencies. \\
                         & $x_n=i\omega_n=i\,2n\pi/\beta$ \\ \\
  {\tt Odd[x1,x2,...]}   & {\tt x1,x2,...} are defined as odd Matsubara
                           frequencies. \\
                         & $x_n=i\omega_n=i(2n+1)\pi/\beta$ \\ \\
\etab
%
%%%%%%%%%%%%%%%%%%%%%%%%%%%%  FIGURES  %%%%%%%%%%%%%%%%%%%%%%%%%%%%%%
%
\newpage
\bfig
  \bc
    \unitlength=1.00mm
    \special{em:linewidth 0.4pt}
    \linethickness{0.4pt}
    \begin{picture}(160.00,55.00)
    \put(0.00,5.00){\framebox(50.00,50.00)[cc]{}}
    \put(55.00,5.00){\framebox(50.00,50.00)[cc]{}}
    \put(25.00,3.00){\makebox(0,0)[cc]{(a)}}
    \put(80.00,3.00){\makebox(0,0)[cc]{(b)}}
    \put(4.94,30.03){\vector(1,0){40.03}}
    \put(24.96,10.01){\vector(0,1){40.03}}
    \put(24.96,44.97){\circle*{2.08}}
    \put(24.96,40.03){\circle*{2.08}}
    \put(24.96,34.96){\circle*{2.08}}
    \put(24.96,30.03){\circle*{2.08}}
    \put(24.96,24.96){\circle*{1.82}}
    \put(24.96,20.02){\circle*{2.08}}
    \put(24.96,14.95){\circle*{1.82}}
    \put(24.96,14.95){\circle{4.16}}
    \put(24.96,20.02){\circle{3.90}}
    \put(24.96,24.96){\circle{4.16}}
    \put(24.96,30.03){\circle{3.90}}
    \put(24.96,34.96){\circle{4.16}}
    \put(24.96,40.03){\circle{3.90}}
    \put(24.96,44.97){\circle{4.16}}
    \put(27.04,44.06){\vector(0,1){0.91}}
    \put(27.04,38.99){\vector(0,1){1.04}}
    \put(27.04,34.05){\vector(0,1){0.91}}
    \put(27.04,28.99){\vector(0,1){1.04}}
    \put(27.04,24.05){\vector(0,1){0.91}}
    \put(27.04,18.98){\vector(0,1){1.04}}
    \put(27.04,14.04){\vector(0,1){0.91}}
    \put(40.03,49.00){\line(0,-1){4.03}}
    \put(40.03,44.97){\line(1,0){4.03}}
    \put(41.98,47.05){\makebox(0,0)[cc]{$z$}}
    \put(30.03,14.95){\makebox(0,0)[cc]{$C$}}
    \put(59.94,30.03){\vector(1,0){40.03}}
    \put(79.96,10.01){\vector(0,1){40.03}}
    \put(79.96,44.97){\circle*{2.08}}
    \put(79.96,40.03){\circle*{2.08}}
    \put(79.96,34.96){\circle*{2.08}}
    \put(79.96,30.03){\circle*{2.08}}
    \put(79.96,24.96){\circle*{1.82}}
    \put(79.96,20.02){\circle*{2.08}}
    \put(79.96,14.95){\circle*{1.82}}
    \put(95.03,49.00){\line(0,-1){4.03}}
    \put(95.03,44.97){\line(1,0){4.03}}
    \put(96.98,47.05){\makebox(0,0)[cc]{$z$}}
    \put(85.03,14.95){\makebox(0,0)[cc]{$C$}}
    \put(81.95,8.06){\vector(0,1){19.95}}
    \put(77.98,52.05){\vector(0,-1){24.04}}
    \put(77.98,28.01){\line(0,-1){19.95}}
    \put(81.95,28.01){\line(0,1){24.04}}
    \put(110.00,5.00){\framebox(50.00,50.00)[cc]{}}
    \put(135.00,3.00){\makebox(0,0)[cc]{(c)}}
    \put(114.94,30.03){\vector(1,0){40.03}}
    \put(134.96,10.01){\vector(0,1){40.03}}
    \put(134.96,44.97){\circle*{2.08}}
    \put(134.96,40.03){\circle*{2.08}}
    \put(134.96,34.96){\circle*{2.08}}
    \put(134.96,30.03){\circle*{2.08}}
    \put(134.96,24.96){\circle*{1.82}}
    \put(134.96,20.02){\circle*{2.08}}
    \put(134.96,14.95){\circle*{1.82}}
    \put(150.03,49.00){\line(0,-1){4.03}}
    \put(150.03,44.97){\line(1,0){4.03}}
    \put(151.98,47.05){\makebox(0,0)[cc]{$z$}}
    \put(137.06,45.04){\vector(0,-1){8.04}}
    \put(132.97,14.99){\vector(0,1){22.02}}
    \put(132.97,37.00){\line(0,1){8.04}}
    \put(137.06,37.00){\line(0,-1){22.02}}
    \put(136.99,30.02){\oval(30.06,30.06)[r]}
    \put(132.97,30.02){\oval(29.93,30.06)[l]}
    \put(152.02,30.94){\vector(0,1){4.08}}
    \put(118.01,37.99){\vector(0,-1){2.97}}
    \put(130.01,12.02){\makebox(0,0)[cc]{$C_1$}}
    \put(140.03,12.02){\makebox(0,0)[cc]{$C_2$}}
    \put(153.01,39.97){\makebox(0,0)[lc]{$\Gamma_2$}}
    \put(117.02,39.97){\makebox(0,0)[rc]{$\Gamma_1$}}
    \end{picture}
  \ec
  \caption{Contours}
  \label{conto}
\efig
\bfig
  \bc
    \unitlength=1.00mm
    \special{em:linewidth 0.4pt}
    \linethickness{0.4pt}
    \begin{picture}(50.00,30.00)
    \put(25.00,15.00){\circle{14.00}}
    \put(0.00,0.00){\framebox(50.00,30.00)[cc]{}}
    \put(5.00,8.00){\line(1,0){40.00}}
    \put(45.00,11.00){\makebox(0,0)[rc]{$p$}}
    \put(5.00,11.00){\makebox(0,0)[lc]{$p$}}
    \put(26.01,21.98){\vector(-1,0){1.02}}
    \put(25.00,26.00){\makebox(0,0)[cc]{$k$}}
    \put(10.03,7.98){\vector(1,0){1.00}}
    \put(38.00,7.98){\vector(1,0){0.99}}
    \end{picture}
  \ec
  \caption{One-loop correction to the self-energy for $\lambda\phi^4$}
  \label{fg:421}
\efig
\bfig
  \epsffile{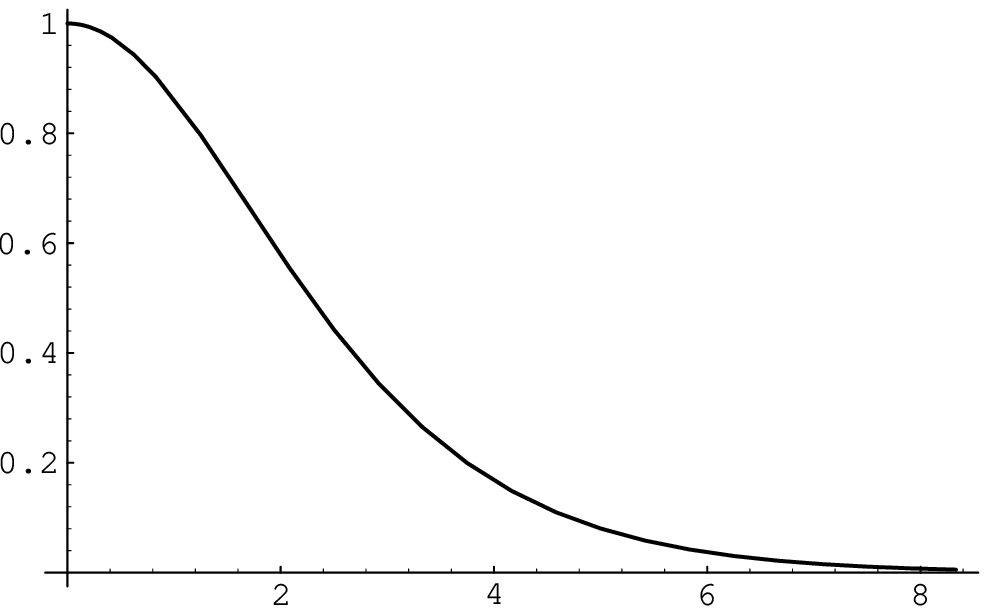}
  \caption{The $g(z)$ function.}
  \label{gfun}
\efig
%
%%%%%%%%%%%%%%%%%%%%%%%%%%%%%%%%%%%%%%%%%%%%%%%%%%%%%%%%%%%%%%%%%%%%%
%
\end{document}